\documentclass[12pt,reqno]{amsart}

\usepackage{amsfonts}
\usepackage{amsmath}
\usepackage{amsthm}
\usepackage{mathrsfs}
\usepackage{amssymb}
\usepackage{stmaryrd}
\usepackage{graphicx}
\usepackage{epstopdf}
\usepackage{textcomp}

\newtheorem{theorem}{Theorem}

\theoremstyle{definition}
\newtheorem{definition}{Definition}

\newtheorem{remark}{Remark}

\newtheorem{example}{Example}

\begin{document}

%
%
%
%
%
%
%
%
%

\title[Minimax perfect stopping rules]{Minimax perfect stopping rules for selling an asset near its ultimate maximum}

\author[D.B. Rokhlin]{Dmitry B. Rokhlin}

\address{%
Institute of Mathematics, Mechanics and Computer Sciences\\
              Southern Federal University\\
Mil'chakova str. 8a\\
344090, Rostov-on-Don\\
Russia
}

\email{rokhlin@math.rsu.ru}
\thanks{The research is supported by Southern Federal University, project 213.01-07-2014/07.}



\begin{abstract}
We study the problem of selling an asset near its ultimate maximum in the minimax setting. The regret-based notion of a perfect stopping time is introduced. A perfect stopping time is uniquely characterized by its optimality properties and has the following form: one should sell the asset if its price deviates from the running maximum by a certain time-dependent quantity. The related selling rule improves any earlier one and cannot be improved by further delay. The results, which are applicable to a quite general price model, are illustrated by several examples.
\end{abstract}

\maketitle

\section{Introduction}
\label{sec:1}
Assume that an agent wants to sell an asset before the maturity date $T$ at a price $X_\tau$, which is as close as possible to the ultimate maximum $X^*_T=\max_{0\le t\le T} X_t$. The asset price is a continuous function $t\mapsto X_t(\omega)$, depending on an unknown outcome $\omega\in\Omega$. A selling rule $\tau(\omega)$ may depend on the price history $\{X_s:s\le\tau(\omega)\}$. For such a rule $\tau$ the difference $X_T^*(\omega)-X_\tau(\omega)$ can be considered as the agent regret that the selling price $X_\tau$ was lower than the maximal price. If the agent is extremely pessimistic, he may try to minimize the value
\begin{equation} \label{1.1}
 \sup_{\omega\in\Omega}\left(X_T^*(\omega)-X_\tau(\omega)\right)
\end{equation}
over all stopping rules $\tau$. However, this approach is somewhat crude.
Such optimal selling rule $\tau^*$ is by no means unique and even a deterministic one (that is, independent of $\omega$)
can be optimal in this sense. Even more importantly, $\tau^*$ need not satisfy Bellman's type optimality principle, as will be clarified below.

To each selling rule we associate the regret over the past, the regret over the future and the overall regret. Based on the latter quantity we introduce the notion of a perfect stopping rule. Considering the family of optimization problems with different initial price histories, we show that under general conditions there is a unique (perfect) stopping rule $\sigma^*$, which is optimal and Pareto optimal with respect to any problem, where it is admissible (Theorem \ref{th:1}). Moreover, $\sigma^*$ can be characterized by the following properties: it improves any earlier stopping rule and cannot be improved by further delay (Theorem \ref{th:2}).

We use two ways to incorporate the regret over the future in the optimization problem. The first one is to consider the maximum of the future price increment. This approach, which is illustrated by Examples \ref{ex:2} and \ref{ex:3}, is very conservative. It is applicable only if the price increments are uniformly bounded. The second approach is to replace the maximum of the future price increment by its $\delta$-quantile in the presence of a probabilistic information. In Example \ref{ex:4} we followed this route
in the case of a Brownian motion (the Bachelier model). Both approaches are captured by the function $\psi$, which can be interpreted as a forecast of the maximal price increment.

The perfect stopping rule has the following simple form: one should sell the asset if its price $X_t$ deviates from the running maximum $X_t^*$ by a certain time-dependent quantity. An optimality of such selling rule (``let profits run but cut losses'') was first justified in \cite{Bawa73} for a discrete time model. This result was inspired by the paper \cite{Pye71}, which studied the case of a divisible asset. The approach of \cite{Pye71,Bawa73} was based on discrete-time specific recurrent dynamic programming formulas. Furthermore, for the problem considered in \cite{Bawa73}, along with the mentioned optimal selling rule, there exists a deterministic (``nonsequential'') selling rule, also minimizing (\ref{1.1}). The notion of a perfect stopping time, introduced below, gives grounds to distinguish between these selling rules and discard the nonsequential one.

In continuous-time probabilistic setting the problem of stopping near the ultimate maximum became popular after the stimulating paper \cite{GraPesShi01} and the preceding talk \cite{Shi02}. For instance, the cases of Brownian motion with drift and geometric Browninan motion were studied thoroughly in \cite{duTPes07,ShiXuZho08,duTPes09,Dai10}. In the latter case the ratios $X_\tau/X_T^*$, $X_T^*/X_\tau$ were considered instead of (\ref{1.1}). Typical optimal stopping rules are determined by the processes $X_t^*-X_t$, $X_t/X_t^*$, or prescribe to sell the asset immediately, or to hold it until the maturity date $T$.

In Section \ref{sec:2} we introduce a perfect stopping time in a general model and present its explicit description. Several illustrative examples are given in Section \ref{sec:3}.

\section{Perfect stopping rule}
\label{sec:2}
Although in this section we do not use any probability measure, the basic terminology comes from probability theory. Possible outcomes (price trajectories) are described by a subset $\Omega$ of the canonical space $C[0,T]$ of continuous functions $\omega$.
Let $X_t:\Omega\mapsto\mathbb R$ be the coordinate mappings: $X_t(\omega)=\omega_t$. For each $t\in [0,T]$, $\omega\in\Omega$ we put $$\mathscr A(t,\omega)=\{\omega'\in\Omega:X_s(\omega')=X_s(\omega),\quad s\in [0,t]\}.$$
The set $\mathscr A(t,\omega)$ contains all outcomes with the same history as $\omega$ up to time $t$. Let us introduce the \emph{regret over the past}:
$$X_t^*(\omega)-X_t(\omega),\quad X_t^*(\omega)=\sup_{s\in [0,t]} X_s(\omega).$$
This quantity corresponds to agent's reflection that he could sell the asset at the price $X_t^*(\omega)$, and now the price is only $X_t(\omega)$.

Similarly, the \emph{regret over the future} is defined as follows:
\begin{equation} \label{2.1}
 \max_{t\le s\le T}X_s(\omega')-X_t(\omega),\quad \omega'\in\mathscr A(t,\omega).
\end{equation}
Since this quantity is unknown unless $t=T$, we will work with its upper estimate or a surrogate of such an estimate.
Consider a function $\psi:[0,T]\times\Omega\mapsto\mathbb [0,\infty)$ with the following properties:
\begin{itemize}
\item[(i)] $\psi(T,\omega)=0$,
\item[(ii)] the function $t\mapsto\psi(t,\omega)$ is continuous and strictly decreasing,
\item[(iii)] $\psi(t,\omega)=\psi(t,\omega')$, $\omega'\in\mathscr A(t,\omega)$.
\end{itemize}

In concrete examples $\psi$ will be taken to be the supremum of (\ref{2.1}) over $\omega'$ (similar to \cite{Bawa73}), or a $\delta$-quantile of (\ref{2.1}), if a probabilistic model is considered. One can also regard $\psi(t,\omega)$ as a \emph{forecast} of the maximal price increment (\ref{2.1}). This interpretation clarifies conditions (i)--(iii). In particular, (iii) means that this forecast can depend only on the available price history.

Assume that the asset is sold at time $u\ge t$. Given the price history $(\omega_s)_{0\le s\le t}$, the \emph{overall regret} can be described by the quantities:
\begin{align*}
\rho(t,\omega;u,\omega') &=\max\left\{X_u^{*}(\omega')-X_u(\omega'),\max_{u\le s\le T}X_s(\omega')-X_u(\omega'))\right\}\\
&=X_T^*(\omega')-X_u(\omega'),\quad \omega'\in\mathscr A(t,\omega).\\
R(t,\omega;u,\omega')&=\max\left\{X_u^{*}(\omega')-X_u(\omega'),\psi(u,\omega')\right\},\quad \omega'\in\mathscr A(t,\omega).
\end{align*}
We call $\rho$ (resp., $R$) the \emph{realized regret} (resp., the \emph{estimated regret}). The realized regret remains unknown until the terminal time $T$. The estimated regret is known at time $u$ and can be incorporated in an optimization problem. In this section we deal only with the estimated regret. The realized regret will be considered in Section \ref{sec:3} (Example \ref{ex:3}).

To formalize agent's goals we need the notion of a stopping time.
\begin{definition}
\label{def:1}
A function $\tau:\Omega\mapsto [0,T]$ is called a \emph{stopping time} if the conditions $\tau(\omega)\le t$, $X_s(\omega')=X_s(\omega)$, $s\le t$ imply that $\tau(\omega')=\tau(\omega)$.
\end{definition}

\begin{remark}
\label{rem:1}
Consider the filtration $\mathscr F_t=\sigma(X_s,s\in[0,t])$, generated by the coordinate mappings. From \cite{DelMey78} (Theorem IV.100 (a)) we know that an $\mathscr F_T$-measurable function $\tau:C[0,T]\mapsto [0,T]$ is a stopping time in the sense of Definition \ref{def:1} if and only if $\{\omega:\tau(\omega)\le s\}\in\mathscr F_s$, $s\in [0,T]$. Thus, our definition of a stopping time coincides with the usual one except of an additional measurability property, which we do not need.
\end{remark}


\begin{remark}
\label{rem:2}
For any stopping time $\tau$ and $\omega\in\Omega$ we have $\tau(\omega')=\tau(\omega)$, $\omega'\in\mathscr A(\tau(\omega),\omega)$, since $X_s(\omega')=X_s(\omega)$, $s\le\tau(\omega)$. Furthermore, for any stopping times $\tau_1$, $\tau_2$ such that $\tau_1(\omega)<\tau_2(\omega)$ we have
$$\tau_1(\omega)=\tau_1(\omega')<\tau_2(\omega'),\quad \omega'\in\mathscr A(\tau_1(\omega),\omega).$$
Indeed, otherwise, $\tau_2(\omega'')\le\tau_1(\omega'')=\tau_1(\omega)$ for some $\omega''\in\mathscr A(\tau_1(\omega),\omega)$. But $X_s(\omega'')=X_s(\omega)$, $s\le \tau_1(\omega)$, and the Definition \ref{def:1} gives a contradiction: $\tau_2(\omega'')=\tau_2(\omega)\le \tau_1(\omega)$.
\end{remark}

Denote by $\mathcal T_t(\omega)$ the set of stopping times $\tau$, satisfying the inequality $\tau(\omega)\ge t$. The condition $\tau\in \mathcal T_t(\omega)$ means that $\tau$ is \emph{admissible} for $\mathscr A(t,\omega)$:
$$\tau(\omega')\ge t,\quad \omega'\in\mathscr A(t,\omega).$$
Given the price history $(\omega_s)_{0\le s\le t}$, the \emph{worst-case estimated regret}, related to $\tau\in \mathcal T_t(\omega)$, is defined as follows:
\begin{align*}
\mathscr R(t,\omega;\tau)&=\sup_{\omega'\in\mathscr A(t,\omega)} R(t,\omega;\tau,\omega'),\\
R(t,\omega;\tau,\omega')&=\max\left\{X_{\tau}^{*}(\omega')-X_\tau(\omega'),\psi(\tau(\omega'),\omega')\right\},\quad X_\tau(\omega)=\omega_{\tau(\omega)}.
\end{align*}

\begin{definition} \label{def:2}
A stopping time $\sigma\in \mathcal T_t(\omega)$ is called \emph{optimal} with respect to $\mathscr A(t,\omega)$ if
$$ \mathscr R(t,\omega;\sigma)\le \mathscr R(t,\omega;\tau),\quad \tau\in\mathcal T_t(\omega).$$
The set of optimal stopping times is denoted by $\mathrm{opt\,}(t,\omega)$.
\end{definition}
\begin{definition} \label{def:3}
A stopping time $\sigma\in \mathcal T_t(\omega)$ is called \emph{Pareto optimal} with respect to $\mathscr A(t,\omega)$ if
there is no $\tau\in\mathcal T_t(\omega)$ such that
\begin{align*}
 R(t,\omega;\tau,\omega') &\le R(t,\omega;\sigma,\omega'),\quad \omega'\in\mathscr A(t,\omega),\\
 R(t,\omega;\tau,\omega'') &< R(t,\omega;\sigma,\omega'')\quad \textrm{for some } \omega''\in\mathscr A(t,\omega).
\end{align*}
The set of Pareto optimal solutions is denoted by $\mathcal P(t,\omega)$.
\end{definition}

\begin{definition} \label{def:4}
We call a stopping time $\sigma$ \emph{perfect} if it satisfies the following \emph{optimality principle}:
$\sigma\in \mathrm{opt\,}(t,\omega)\cap\mathcal P(t,\omega)$ for all $(\omega,t)$ such that $\sigma\in\mathcal T_t(\omega).$
\end{definition}
That is, $\sigma$ is perfect if it is \emph{optimal and Pareto optimal} with respect to $\mathscr A(t,\omega)$ \emph{whenever it is admissible} for $\mathscr A(t,\omega)$.
\begin{theorem} \label{th:1}
Assume that for any $(t,\omega)\in [0,T]\times\Omega$ there exists $\widehat\omega\in\mathscr A(t,\omega)$ such that $X_u(\widehat\omega)<X_t(\omega)$, $u>t$. Then the unique perfect stopping time is given by the formula
$$\sigma^*(\omega)=\inf\{s\ge 0:(X_s^*-X_s)(\omega)\ge \psi(s,\omega)\}.$$
\end{theorem}
\emph{Proof.} Since $(X_0^*-X_0)(\omega)<\psi(0,\omega)$ and $(X_T^*-X_T)(\omega)\ge \psi(T,\omega)$, the value $\sigma^*(\omega)$ is uniquely defined and satisfies the equality
\begin{equation} \label{2.2}
X_{\sigma^*}^*-X_{\sigma^*}=\psi(\sigma^*,\omega),\quad \omega\in\Omega.
\end{equation}

Assume that $\sigma^*(\omega)\ge t$ and take $\tau\in\mathcal T_t(\omega)$. If $\tau(\omega')<\sigma^*(\omega')$ for some $\omega'\in\mathscr A(t,\omega)$, then
\begin{align} \label{2.3}
\mathscr R(t,\omega;\tau) & \ge R(t,\omega;\tau,\omega')\ge\psi(\tau(\omega'),\omega')\nonumber\\
&>\psi(\sigma^*(\omega'),\omega')=R(t,\omega;\sigma^*,\omega'),
\end{align}
where the strict inequality follows from the property (ii) of $\psi$ and the last equality is implied by (\ref{2.2}). Furthermore, if $\tau(\omega')>\sigma^*(\omega')$, then take $\widehat\omega\in\mathscr A(\sigma^*(\omega'),\omega')$ such that
\begin{equation} \label{2.4}
X_u(\widehat\omega)<X_{\sigma^*}(\omega'),\quad u>\sigma^*(\omega').
\end{equation}
We have
\begin{align} \label{2.5}
\mathscr R(t,\omega;\tau) &\ge R(t,\omega;\tau,\widehat\omega)\ge (X_\tau^*-X_\tau)(\widehat\omega) =X_{\sigma^*}^*(\omega')-X_\tau(\widehat\omega)\nonumber\\
&>X_{\sigma^*}^*(\omega')-X_{\sigma^*}(\omega')=R(t,\omega;\sigma^*,\widehat\omega),
\end{align}
where the strict inequality and the last equality follow from (\ref{2.4}) and (\ref{2.2}) respectively. Finally, if $\tau(\omega')=\sigma^*(\omega')$, then
\begin{equation} \label{2.6}
\mathscr R(t,\omega;\tau) \ge R(t,\omega;\tau,\omega')=R(t,\omega;\sigma^*,\omega').
\end{equation}

The relations (\ref{2.3}), (\ref{2.5}), (\ref{2.6}) imply that $\mathscr R(t,\omega;\tau)\ge \mathscr R(t,\omega;\sigma^*)$, and if $\tau\neq\sigma^*$, then there exists a point $\omega''\in\mathscr A(t,\omega)$ such that
$$R(t,\omega;\tau,\omega'')>R(t,\omega;\sigma^*,\omega'').$$
Thus, $\sigma^*\in \mathrm{opt\,}(t,\omega)\cap \mathcal P(t,\omega)$.

Now let $\tau$ be any stopping time. If $\tau(\omega)<\sigma^*(\omega)$ for some $\omega\in\Omega$, then $\tau(\omega')=\tau(\omega)<\sigma^*(\omega')$ for all $\omega'\in\mathscr A(t,\omega)$, where $t=\tau(\omega)$. Hence, the inequality (\ref{2.3}):
$$R(\tau(\omega),\omega;\tau,\omega')>R(\tau(\omega),\omega;\sigma^*,\omega'),\quad \omega'\in\mathscr A(\tau(\omega),\omega)$$
implies that $\tau$ is not Pareto optimal for $\mathscr A(\tau(\omega),\omega)$.

Furthermore, if $\tau(\omega)>\sigma^*(\omega)$, then $\tau(\omega')>\sigma^*(\omega')=\sigma^*(\omega)$ for all $\omega'\in\mathscr A(t,\omega)$, where $t=\sigma^*(\omega)$. Hence, the inequality (\ref{2.5}) implies that $\tau$ is not optimal with respect to $\mathscr A(\sigma^*(\omega),\omega)$:
$$\mathscr R(\sigma^*(\omega),\omega;\tau)>\mathscr R(\sigma^*(\omega),\omega;\sigma^*),$$
where we used the fact that $\sigma^*(\omega')$ and $R(t,\omega;\sigma^*,\omega')$ do not depend on $\omega'\in\mathscr A(\sigma^*(\omega),\omega)$.
\qed

The following assertion gives more intuition on perfect stopping times.
\begin{theorem} \label{th:2}
Under assumption of Theorem \ref{th:1} a stopping time $\sigma^*$ is perfect if and only if for any stopping time $\tau$ and any $\omega\in\Omega$ the following is true:
\begin{itemize}
\item[(A)] if $\tau(\omega)>\sigma^*(\omega)$ then
$$  R(\sigma^*(\omega),\omega;\tau,\omega')> R(\sigma^*(\omega),\omega;\sigma^*,\omega')\quad \textrm{for some } \omega'\in\mathscr A(\sigma^*(\omega),\omega),$$
\item[(B)] if $\tau(\omega)<\sigma^*(\omega)$ then
$$ R(\tau(\omega),\omega;\tau,\omega')> R(\tau(\omega),\omega;\sigma^*,\omega')\quad \textrm{for all } \omega'\in\mathscr A(\tau(\omega),\omega).$$
\end{itemize}
\end{theorem}

Condition (A) (``after'') means that the estimated regret can become larger if the asset is not sold at time $\sigma^*$.
Condition (B) (``before'') means that it is not rational to sell the asset before a perfect stopping time $\sigma^*$, since the estimated regret can be reduced by waiting until $\sigma^*$.

\emph{Proof of Theorem \ref{th:2}.} A stopping time $\sigma^*$, satisfying conditions (A), (B) is unique. Indeed, let $\sigma_1$, $\sigma_2$ be such stopping times. If $\sigma_1(\omega)<\sigma_2(\omega)$, then
$$ R(\sigma_1(\omega),\omega;\sigma_1,\omega')>R(\sigma_1(\omega),\omega;\sigma_2,\omega'),\quad \omega'\in\mathscr A(\sigma_1(\omega),\omega),$$
since $\sigma_2$ satisfies (B), and
$$ R(\sigma_1(\omega),\omega;\sigma_2,\omega'')>R(\sigma_1(\omega),\omega;\sigma_1,\omega'')\quad \textrm{for some }\omega''\in\mathscr A(\sigma_1(\omega),\omega),$$
since $\sigma_1$ satisfies (A). This contradiction indicates that $\sigma_1\ge\sigma_2$. By symmetry, $\sigma_1=\sigma_2$.

Furthermore, the perfect stopping time $\sigma^*$ satisfies (A), (B), as was shown in the course of the proof of Theorem \ref{th:1}: apply (\ref{2.5}) with $t=\sigma^*(\omega)$ and (\ref{2.3}) with $t=\tau(\omega)$.
\qed

\section{Examples}
In this section we present several examples, illustrating the above notions and results. Example \ref{ex:1} indicates that a stopping time can be optimal but not Pareto optimal, or vice versa. Example \ref{ex:2} considers a continuous time analogue of the model of \cite{Bawa73}. In Example \ref{ex:3} we consider a price process with piecewise-linear trajectories, which can change their movement direction at the jump times of a Poisson process. Here we compare the expected realized regret of the perfect and deterministic stopping times. Finally, in Example \ref{ex:4} we analyze the classical Bachelier model, using the quantile function to estimate the future regret.
\label{sec:3}
\begin{example} \label{ex:1}
Let $\Omega=C[0,T]$, and assume that $\psi$ does not depend on $\omega$. We claim that $\tau_0=0$ is optimal, but not Pareto optimal with respect to $\mathscr A(0,\omega)$, and $\tau_T=T$ is Pareto optimal, but not optimal with respect to $\mathscr A(t,\omega)$ for any $t<T$.

Clearly, $\mathscr R(0,\omega;\tau_0)=\psi(0)$. Take $\omega^n=\omega_0-ns$. By (\ref{2.2}) for the perfect stopping time $\sigma^*$ we have  $$\mathscr R(0,\omega;\sigma^*)\ge R(0,\omega;\sigma^*,\omega^n)=\psi(\sigma^*(\omega^n)).$$ But,
$$ \sigma^*(\omega^n)=\inf\{s\ge 0:X_s^*(\omega^n)-X(\omega^n)\ge\psi(s)\}=\inf\{s\ge 0:ns\ge\psi(s)\}\to 0,$$
as $n\to\infty$. Hence, $\mathscr R(0,\omega;\sigma^*)=\mathscr R(0,\omega;\tau_0)=\psi(0)$ and $\tau_0\in\mathrm{opt\,}(0,\omega)$ along with $\sigma^*$. However, $\tau_0$ is not Pareto optimal with respect to $\mathscr A(0,\omega)$, as can be seen from the property (B) of Theorem \ref{th:2}, where $\tau$ is changed to $\tau_0$.

As for $\tau_T$, it is not optimal with respect to $\mathscr A(t,\omega)$, $t<T$, since
$$\mathscr R(t,\omega;\tau_T)=\sup_{\omega'\in\mathscr A(t,\omega)}(X_T^*-X_T)(\omega')=+\infty.$$
The same assertion is true for any deterministic stopping time $\tau=u$ and $t<u$.

To prove that $\tau_T$ is Pareto optimal assume that $t\le\tau(\omega)=u<T$ and put
$$\omega'_s=\begin{cases}
\omega_s,& s\le u\\
\omega_u+s-u,& s\ge u.
\end{cases}$$
If $X_u^*(\omega')>X_u(\omega')$, then
\begin{align} \label{3.1}
R(t,\omega;\tau_T,\omega')&=(X_T^*-X_T)(\omega')<\max\{(X_u^*-X_u)(\omega'),\psi(u)\}\nonumber\\
&=\max\{(X_\tau^*-X_\tau)(\omega'),\psi(\tau(\omega'))\}=R(t,\omega;\tau,\omega').
\end{align}
If $X_u^*(\omega')=X_u(\omega')$, then $X_T^*(\omega')=X_T(\omega')$ and
\begin{equation} \label{3.2}
0=R(t,\omega;\tau_T,\omega')<\psi(u)=R(t,\omega;\tau,\omega').
\end{equation}
The inequalities (\ref{3.1}), (\ref{3.2}) show that $\tau_T\in\mathcal P(t,\omega)$.
\end{example}

\begin{example} \label{ex:2}
Consider the set $\Omega$ of $\omega\in C[0,T]$ such that
\begin{equation} \label{3.3}
 -L_1\cdot (t-s)\le X_t(\omega)-X_s(\omega)\le L_2\cdot(t-s),\quad 0\le s<t\le T
\end{equation}
with some constants $L_1,L_2>0$. In particular, $\omega\in\Omega$ are assumed to be uniformly Lipschitz continuous. Note, that any piecewise linear function
$$ \omega_t=\omega_{t_i}+\frac{t-t_i}{t_{i+1}-t_i}(\omega_{t_{i+1}}-\omega_{t_i}),\quad t\in [t_i,t_{i+1}],$$
where $0=t_0<t_1<\dots<t_n=T$ and
$-L_1\le(\omega_{t_{i+1}}-\omega_{t_i})/(t_{i+1}-t_i)\le L_2,$
belongs to $\Omega$.

Let us assume that the forecasted maximal price increment $\psi$ coincides with the maximum itself:
$$\psi(t,\omega)=\sup_{\omega'\in\mathscr A(t,\omega)}\sup_{s\in [t,T]} X_s(\omega')-X_t(\omega).$$
Clearly, $\psi\le L_2\cdot(T-t)$. Moreover, as
$$ \omega''_s=\begin{cases}
\omega_s,& s\le t\\
\omega_t+L_2\cdot (s-t),& s\ge t
\end{cases}$$
belongs to $\mathscr A(t,\omega)$, it follows that $\psi=L_2\cdot(T-t)$.

The perfect stopping time is defined by
\begin{equation} \label{3.4}
\sigma^*(\omega)=\inf\{s\ge 0: (X_s^*-X_s)(\omega)\ge L_2\cdot (T-s)\}.
\end{equation}
Note, that it depends only on one parameter $L_2$, which shows how fast the price can go upwards.

Assume that $\sigma^*(\omega)\ge t$. Let $\widetilde\tau:\Omega\mapsto [t,\sigma^*]$ be a random variable such that $$X_{\widetilde\tau}=\max_{t\le s\le\sigma^*} X_s.$$
Note, that $\widetilde\tau$ need not be a stopping time. By the definition of $\sigma^*$ and the left inequality (\ref{3.3}) we get
\begin{align*}
L_2 (T-\sigma^*)&=\psi(\sigma^*)=X^*_{\sigma^*}-X_{\sigma^*}=\max\{X_t^*,\max_{t\le s\le\sigma^*}X_s\}-X_{\sigma^*}\\
&=\max\{X_t^*-X_{\sigma^*},X_{\widetilde\tau}-X_{\sigma^*}\}\\
&=\max\{X_t^*-X_t-(X_{\sigma^*}-X_t),-(X_{\sigma^*}-X_{\widetilde\tau})\}\\
&\le\max\{X_t^*-X_t+L_1(\sigma^*-t),L_1(\sigma^*-\widetilde\tau)\}\\
&\le X_t^*-X_t+L_1(\sigma^*-t).
\end{align*}
Thus,
\begin{equation} \label{3.5}
\sigma^*\ge\frac{L_1}{L_1+L_2} t+ \frac{L_2}{L_1+L_2} T-\frac{X_t^*-X_t}{L_1+L_2}.
\end{equation}
Moreover, for $\overline\omega_s=\omega_t-L_1 (s-t)$, $s\ge t$ we have
$$ (X_{\sigma^*}^*-X_{\sigma^*})(\overline\omega)=X_t^*(\omega)-X_t(\omega)+L_1(\sigma^*(\overline\omega)-t)
=L_2\cdot(T-\sigma^*(\overline\omega)).$$
Hence, $\sigma^*(\overline\omega)$ coincides with the lower bound (\ref{3.5}).
It follows that the optimal worst-case estimated regret ${\mathscr R}(t,\omega;\sigma^*)$ coincides with some convex combination of the regret over the past $X_t^*-X_t$ and the estimated regret over the future $L_2(T-t)$:
\begin{align} \label{3.6}
\inf_{\tau\in\mathcal T_t}{\mathscr R}(t,\omega;\tau) &= {\mathscr R}(t,\omega;\sigma^*)=\sup_{\omega'\in\mathscr A(t,\omega)}L_2(T-\sigma^*(\omega'))=L_2(T-\sigma^*(\overline\omega))\nonumber\\
&=\frac{L_2}{L_1+L_2}(X_t^*-X_t)+\frac{L_1}{L_1+L_2}L_2(T-t).
\end{align}

Let $t=0$. Note, that along with $\sigma^*$, the deterministic stopping time
$$\widehat\tau=\frac{L_2 T}{L_1+L_2}$$
is optimal with respect to $\mathscr A(0,\omega)$. Indeed, since $\widehat\tau\le\sigma^*$ (see (\ref{3.5})), we have
$$ X^*_{\widehat\tau}-X_{\widehat\tau}\le\psi(\widehat\tau)=L_2\left(T-\widehat\tau\right)=\frac{L_1 L_2}{L_1+L_2}T.$$
It follows that
$${\mathscr R}(0,\omega;\widehat\tau)=\sup_{\omega'\in\mathscr A(0,\omega)}\max\{(X^*_{\widehat\tau}-X_{\widehat\tau})(\omega'),\psi(\widehat\tau)\}\le \frac{L_1 L_2}{L_1+L_2}T={\mathscr R}(0,\omega;\sigma^*).$$
But the strict inequality is impossible, since $\sigma^*$ is optimal. Optimal stopping times quite similar to $\sigma^*$, $\widehat\tau$ appeared in \cite{Bawa73}.

Let us mention the following clear advantage of the perfect stopping time $\sigma^*$ over $\widehat\tau$. Comparing the expression
$$ \mathscr R(\widehat\tau,\omega;\widehat\tau)=\max\{X^*_{\widehat\tau}-X_{\widehat\tau},L_2(T-\widehat\tau)\}$$
with (\ref{3.6}) we conclude that $\widehat\tau$ is not optimal with respect to $\mathscr A(\widehat\tau,\omega)$ unless
\begin{equation} \label{3.7}
X^*_{\widehat\tau}-X_{\widehat\tau}=L_2(T-\widehat\tau).
\end{equation}
Hence, although at time $t=0$ formally both $\sigma^*(\omega)$, $\widehat\tau$ are optimal, the agent, who observes the price dynamics $X_s(\omega)$, $0\le s\le t$, at time $\widehat\tau$ realizes that it is not optimal to sell the asset 	
with the exception of a rather special situation, described by (\ref{3.7}). Similarly, at any time moment $t\in (0,\widehat\tau)$ typically $\widehat\tau$ ceases to be optimal with respect to $\mathscr A(t,\omega)$. This violation of ''Bellman's optimality principle'' means that $\widehat\tau$ will be quite rarely used by a rational agent. Recall also that $\widehat\tau$ is not Pareto optimal with respect to $\mathscr A(\widehat\tau,\omega)$ if $\widehat\tau<\sigma^*(\omega)$: see condition (B) of Theorem \ref{th:2}.
\end{example}

\begin{example} \label{ex:3}
Assume that $\Omega$ and $\psi$ are the same as in Example \ref{ex:2}, and $L_1=L_2=1$. Consider a probability space, carrying a Poisson process $(N_t)_{t\ge 0}$ with intensity $\lambda$ and a sequence of i.i.d. random variables $(\alpha_i)_{i=1}^\infty$:
$$ \mathsf P(\alpha_i=1)=p\in (0,1),\quad \mathsf P(\alpha_i=-1)=q=1-p,$$
which are independent from $N$. The asset price is modeled by the continuous piecewise-linear process
$$ X_t=X_{\tau_i}+\alpha_{i+1} (t-\tau_i),\quad t\in [\tau_i,\tau_{i+1}), \quad i\ge 0,$$
where $\tau_0=0$, $X_0=x$ and $\tau_i$, $i\ge 1$ are the jump times of $N$.

For any stopping time $\tau$ denote by
\begin{align*}
\widehat{\mathcal E}(\tau)&=\mathsf E \rho(0,\omega;\tau,\cdot)=\mathsf E (X_T^*-X_\tau),\\
\mathcal E(\tau) &=\mathsf E R(0,\omega;\tau,\cdot)=\mathsf E\max\left\{X_{\tau}^{*}-X_\tau,\psi(\tau)\right\}
\end{align*}
the \emph{expected realized regret} and \emph{expected estimated regret} respectively.
To obtain some information, concerning the probabilistic properties of the perfect stopping time (\ref{3.4}), let us compare $\widehat{\mathcal E}(\sigma^*)$, $\mathcal E(\sigma^*)$ with the same values, related to deterministic stopping times $\tau_u=u$.

Let the intensity $\lambda$ be very small, so that that there are no jumps on $[0,T]$ with high probability. Then with probability close to $1$ the set $\mathscr A(0,\omega)$ contains only two trajectories:
$$ \omega^1_s=\omega_0+s,\quad \omega^2_s=\omega_0-s,\quad s\in [0,T].$$
Note, that by observing a trajectory for a short period of time it is possible to conclude, if the price will permanently go up or down. Thus, it is possible to obtain an arbitrary small regret by stopping after this period of time or waiting until $T$. Our aim, however, is only to compare the expected regret of the perfect stopping time and a deterministic one.

We have
\begin{align} \label{3.8}
 \mathcal E(\tau_u)&\approx p\max\{0,T-u\}+q\max\{u,T-u\}\nonumber\\
 &=\max\{pT+(q-p)u,T-u\}\ge\begin{cases}
 qT,& p\ge 1/2\\
 T/2,& p<1/2,
 \end{cases}
\end{align}
where the lower bound is attained at $\overline u=T$ for $p\ge 1/2$ and at $\overline u=T/2$ for $p<1/2$.
Furthermore,
\begin{equation} \label{3.9}
\mathcal E(\sigma^*)=\mathsf E\psi(\sigma^*)=T-\mathsf E\sigma^*\approx qT/2.
\end{equation}
since $\sigma^*(\omega^1)=T$, $\sigma^*(\omega^2)=T/2$. On the basis of (\ref{3.8}), (\ref{3.9}) it is reasonable to expect that
$\mathcal E(\sigma^*)<\mathcal E(\tau_u)$ for any $u$ if the intensity $\lambda$ is sufficiently small.

For the expected realized regret this conclusion should be refined. We have
$$ X_T^*(\omega^1)-X_u(\omega^1)=T-u,\quad X_T^*(\omega^2)-X_u(\omega^2)=u.$$
Thus,
$$\widehat{\mathcal E}(\sigma^*)=\mathsf E(X_T^*-X_{\sigma^*})\approx p (T-\sigma^*(\omega^1))+q\sigma^*(\omega^2)= qT/2,$$
$$ \widehat{\mathcal E}(\tau_u)\approx p (T-u)+q u=pT +(q-p)u\ge \begin{cases}
qT,& p\ge 1/2\\
pT,& p<1/2,
\end{cases}
$$
where the lower bound is attained at $u=T$ for $p\ge 1/2$ and at $u=0$ for $p<1/2$. So, we expect the inequality
$$\widehat{\mathcal E}(\sigma^*)<\widehat{\mathcal E}(\tau_u)$$
to be true for small $\lambda$ if $q/2<p$, that is, $p>1/3$.

These expectations are confirmed by numerical experiments. To estimate $\widehat{\mathcal E}(\tau)$ ee used the Monte Carlo method (and $\mathsf R$ software). Note, that the variances of $X_\tau^*-X_\tau$, $\sigma^*$ does not exceed $T^2$. For $T=1$ we used $N=10^6$ samples of these random variables. A standard asymptotic analysis based on the normal approximation (see, e.g., \cite[Chapter 4]{Wang12}) allows to conclude that the length of 0.99 confidence interval does not exceed 0.006.

The results for $p=1/2$ and different $\lambda$ are collected in Table \ref{tab:1}. In this case the values $\mathcal E(\tau_u)$ do not depend on $u$. We see that for small values of $\lambda$ the expected realized regret of the perfect stopping time is essentially lower than that of a deterministic stopping time. This advantage diminishes with the growth of $\lambda$. The value $\mathsf E\sigma^*$ for large $\lambda$ is close to $T$.

In Table \ref{tab:2} the value $\lambda=10$ is fixed. For large $p$ the expected realized regret of $\sigma^*$ is smaller than that of any deterministic stopping time. For small $p$ an immediate selling $(\tau_0=0$) has much smaller expected realized regret than $\sigma^*$. Note, that $\sigma^*$ cannot be smaller than $T/2$ (see Example \ref{ex:2}). Intuitively, this is due to the fact that the estimate $\psi$ of the future regret is very conservative.

\begin{table}
\caption{The expected realized regret for $p=1/2$, $T=1.$}
\label{tab:1}       
\begin{tabular}{llllll}
\hline\noalign{\smallskip}
$\lambda$ & $\mathsf E\sigma^*$ & $\widehat{\mathcal E}(\sigma^*)$ & $\widehat{\mathcal E}(u)$ \\
\noalign{\smallskip}\hline\noalign{\smallskip}
0.1 & 0.75& 0.25 & 0.49 \\
1   & 0.75& 0.25 & 0.45 \\
10  & 0.8 & 0.2 & 0.27\\
50  & 0.88 & 0.12 & 0.14\\
100 & 0.9 & 0.09& 0.1 \\
1000 & 0.97 & 0.03 & 0.03\\
\noalign{\smallskip}\hline
\end{tabular}
\end{table}

\begin{table}
\caption{The expected realized regret for $\lambda=10$, $T=1.$}
\label{tab:2}       
\begin{tabular}{llllll}
\hline\noalign{\smallskip}
$p$ & $\mathsf E\sigma^*$ & $\widehat{\mathcal E}(\sigma^*)$ & $\widehat{\mathcal E}(0)$ & $\widehat{\mathcal E}(T/2)$ & $\widehat{\mathcal E}(T)$  \\
\noalign{\smallskip}\hline\noalign{\smallskip}
0.2 & 0.61 & 0.39 & 0.06 &  0.36 & 0.66\\
0.4 & 0.74 & 0.26 & 0.18 &  0.28 & 0.38\\
0.6 & 0.86 & 0.14 & 0.38 &  0.28 & 0.18\\
0.8 & 0.95 & 0.05 & 0.66 &  0.36 & 0.06 \\

\noalign{\smallskip}\hline
\end{tabular}
\end{table}

\end{example}

\begin{example} \label{ex:4}
Let $\mathsf P$ be a probability measure on $(C[0,T],\mathscr F_T)$ such that the coordinate process $X$ follows the Bachelier model under $\mathsf P$:
$$X_t=x+\sigma W_t. $$
Here $W$ is a standard Brownian motion, started at $0$, and $\sigma>0$ is a volatility constant. For this model the regret over the future (\ref{2.1}) is unbounded. So we use quantile function
$$ \psi(t,\omega)=\inf\left\{z:\mathsf P\left(\omega':\max_{t\le s\le T} X_s(\omega')-X_t(\omega)\le z\right)\ge\delta\right\}.$$
Utilizing this function in the estimated regret, the agent admits that the realized future regret, that is, the deviation of the ultimate maximum from the selling price, can exceed $\psi$ with probability $1-\delta$.

Using the law of the running maximum $W^*$ of the Brownian motion $W$ (see, e.g., \cite[Proposition 3.1.3.1]{JeaYorChe09}):
\begin{align*}
\mathsf P\left(W^*_u\le z\right) &=\mathsf P(|W_u|\le z)=2\mathsf P(W_u\in [0,z])=2\Phi\left(\frac{z}{\sqrt u}\right)-1,\\
 \Phi(y)&=\sqrt\frac{1}{2\pi }\int_{-\infty}^y \exp\left(-\frac{x^2}{2}\right)\,dx,
\end{align*}
we infer that $\psi$ solves the equation
$$ 2\Phi\left(\frac{\psi}{\sigma\sqrt{T-t}}\right)-1=\delta.$$
Thus, $\psi=\sigma\Phi^{-1}((1+\delta)/2)\sqrt{T-t}$. By Theorem \ref{th:1} the perfect stopping time is defined by the formula
\begin{align} \label{3.10}
\sigma^*(\omega) &=\inf\{s\ge 0:W_s^*-W_s\ge c_\delta\sqrt{T-s}\}\ \quad \mathsf P\textrm{-a.s.},\\ c_\delta&=\sigma\Phi^{-1}((\delta+1)/2).\nonumber
\end{align}
Note, that given a price history, the agent will keep the asset for a longer time period when the market is more volatile. This is due to the fact that his estimated future regret will be higher.

In contrast to Example \ref{ex:3}, here it does not make sense to look at the expected realized regret $\widehat{\mathcal E}(\tau)$, since $W$ is a martingale and $\widehat{\mathcal E}(\tau)$ does not depend on $\tau$ by the Doob optional sampling theorem. It appears, however, that $\sigma^*$ is optimal with respect to another criteria.

In \cite{Ped03} it was proved that the stopping time of the form
$$ \tau^*=\inf\{s\ge 0:W_s^*-W_s\ge z_q\sqrt{T-s}\}$$
is optimal for the $q$-mean objective function
\begin{equation} \label{3.11}
\mathsf E(W_T^*-W_\tau)^q\to\min_\tau,\quad q>1,
\end{equation}
where minimization is performed over all stopping times $\tau$. The number $z_q$ is the unique positive root of the equation
\begin{equation} \label{3.12}
\frac{H'(z)}{H(z)}+z=(1+q)z\frac{M(\frac{3+q}{2},\frac{3}{2},\frac{1}{2}z^2)}{M(\frac{1+q}{2},\frac{1}{2},\frac{1}{2}z^2)},
\end{equation}
where $H(z)=z^q+2\int_{z^q}^\infty(1-\Phi(u^{1/q}))\,du$, $z\ge 0$ and
$$ M(a,b,z)=1+\frac{a}{b}z+\frac{1}{2!}\frac{a(a+1)}{b(b+1)}z^2+\cdots$$
is the Kummer confluent hypergeometric function (see \cite[Chapter 13]{AbrSte72}). The case $q=2$ was previously considered in \cite{GraPesShi01}.

Put $\sigma=1$. Given $q>1$, we infer that the perfect stopping time (\ref{3.10}) is $q$-mean optimal in the sense of (\ref{3.11}) for the special value of $\delta$. The correspondence between some values of $q$ and $\delta$ is presented in Table \ref{tab:3} (the equation (\ref{3.12}) was solved by the means of the $\mathsf R$ software).
For instance, if $\psi$ is chosen to be 0.95-quantile of the maximum future price increment, then the perfect stopping time $\sigma^*$ is $q$-mean optimal with $q=10$. The values of $z_q$ in the first 3 rows coincide with those of \cite{Ped03}.

\begin{table}
\caption{The values of $q$, $z_q$ and the correspondent values of $\delta=2\Phi(z_q)-1$ for $\sigma=1$.}
\label{tab:3}       
\begin{tabular}{lll}
\hline\noalign{\smallskip}
$q$ & $z_q=c_\delta$ & $\delta$ \\
\noalign{\smallskip}\hline\noalign{\smallskip}
1.1 & 1.03 & 0.7 \\
2 & 1.12 & 0.74 \\
4 & 1.35 & 0.82 \\
6 & 1.57 & 0.88 \\
8 & 1.77 & 0.92 \\
10 &  1.96 & 0.95\\
\noalign{\smallskip}\hline
\end{tabular}
\end{table}
\end{example}

\section{Concluding remarks}
The relations between minimax and probabilistic optimality properties of stopping times, touched in Examples \ref{ex:3}, \ref{ex:4}, may deserve further study. We also mention  that for a strictly positive price process $X$ the above theory can be transferred to the ratio performance criterion $X_T^*/X_\tau$, which depends only on relative values of the asset price.


\begin{thebibliography}{10}

\bibitem{AbrSte72}
M.~Abramowicz and I.A. Stegun, editors.
\newblock {\em Handbook of mathematical functions with formulas, graphs, and
  mathematical tables}, volume~55 of {\em NBS Appl. Math. Series}.
\newblock Washington, 1972.

\bibitem{Bawa73}
V.~S. Bawa.
\newblock Minimax policies for selling a nondivisible asset.
\newblock {\em Manage. Sci.}, 19(7):760--762, 1973.

\bibitem{Dai10}
M.~Dai, H.~Jin, Y.~Zhong, and X.~Y. Zhou.
\newblock Buy low and sell high.
\newblock In {\em Contemporary quantitative finance}, pages 317--333. Springer,
  2010.

\bibitem{DelMey78}
C.~Dellacherie and P.-A. Meyer.
\newblock {\em Probabilities and potential}.
\newblock Amsterdam, North-Holland, 1978.

\bibitem{duTPes07}
J.~du~Toit and G.~Peskir.
\newblock The trap of complacency in predicting the maximum.
\newblock {\em Ann. Probab.}, 35(1):340--365, 2007.

\bibitem{duTPes09}
J.~du~Toit and G.~Peskir.
\newblock Selling a stock at the ultimate maximum.
\newblock {\em Ann. Appl. Probab.}, 19(3):983--1014, 2009.

\bibitem{GraPesShi01}
S.~E. Graversen, G.~Peskir, and A.~N. Shiryaev.
\newblock Stopping {B}rownian motion without anticipation as close as possible
  to its ultimate maximum.
\newblock {\em Theory Probab. Appl.}, 45(1):125--136, 2001.

\bibitem{EvaHenHob08}
Evans J., V.~Henderson, and D.~Hobson.
\newblock Optimal timing for an indivisible asset sale.
\newblock {\em Math. Finance}, 18(4):545--567, 2008.

\bibitem{JeaYorChe09}
M.~Jeanblanc, M.~Yor, and M.~Chesney.
\newblock {\em Mathematical methods for financial markets}.
\newblock Springer, London, 2009.

\bibitem{Ped03}
J.L. Pedersen.
\newblock Optimal prediction of the ultimate maximum of {B}rownian motion.
\newblock {\em Stoch. Stoch. Rep.}, 75(4):205--219, 2003.

\bibitem{Pye71}
G.~Pye.
\newblock Minimax policies for selling an asset and dollar averaging.
\newblock {\em Manage. Sci.}, 17(7):379--393, 1971.

\bibitem{ShiXuZho08}
A.~Shiryaev, Z.~Xu, and X.~Y. Zhou.
\newblock Thou shalt buy and hold.
\newblock {\em Quant. Finance}, 8(8):765--776, 2008.

\bibitem{Shi02}
A.~N. Shiryaev.
\newblock Quickest detection problems in the technical analysis of the
  financial data.
\newblock In {\em Proc. Math. Finance Bachelier Congress (Paris, 2000)}, pages
  487--521. Springer, 2002.

\bibitem{Wang12}
H.~Wang.
\newblock {\em Monte Carlo Simulation with Applications to Finance}.
\newblock CRC Press, Boca Raton, 2012.

\end{thebibliography}

\end{document}